# ON THE SURFACE WAVES IN THE SOLAR PHOTOSPHERE

**M.B. KERIMBEKOV**

*Shamakhy Astrophysical Observatory*

*Azerbaijan Academy of Sciences*
*Observatory, settl. Y. Mamedaliyev, Shamakhy*

The regular structures similar to the chains and clusters on the Solar Photosphere are investigated and mechanism these origin are proposed.

Scrutinising direct pictures, obtained by the method of cinematography, one may find the structures similar to the chains and clusters. The same configurations may be observed in high. Photosphere in the line: $H_\alpha$-1 Å, etc. The configurations differ from each other due to their shapes and sizes.

Their considerable length may be due to two mechanisms of the origin (waves and Percolation processes). In this paper we'll discuss the first one. Recently, it was indicated that at the frontiers of supergranulaes there were small-scale spin waves ($\lambda \leq 10^7$) cm [1]. The explanation of this phenomenon is very difficult from classical point of view. That is why we to turn to the quantum mechanical interpretation.

According to the last one solar plasma partly degenerates at the level under the surface where the $n$ is equal $10^{17} \div 10^{21}$ cm charged particles per cm$^3$ and $T$ is $\cong 10^4$ K. Here the distance between particles are $r = 10 \div 100\ \lambda_B$, where $\lambda_B = \dfrac{\hbar}{p} = 10^{-6} \div 10^{-7}$, ($r = \lambda_B$) and orbites are overlapping. It leads to the origin of collective processes in the continuum [2] and to the arrangement of magnetic momenta $M_{orb.}$ in "domains" The value of $M_0$ may be of order:

$$|M| = 4\pi \frac{eVr}{c} n\mu$$

$\mu$ is magnetic permeability.

The "domains", located previously at the same levels under the surface came up to the photosphere due to the mechanism of E. Parker [3]. Analyses of dispersion relationship show that the shortest part of spin wave is most energetic ($\varepsilon \sim \vartheta_{group} \sim \lambda^{-1};\ \lambda_s^{min} \equiv \lambda_{deby}$). But what is the longest wavelength? To answer on this question let us consider the electron - ion interaction leading to the broadening of the spin wave spectra. The potential of the interaction is $V = \dfrac{M_e M_i}{r_0^3}$; where $r_0$ is the mean distance between the particles. And $M_{e,i} = \dfrac{er_0}{c}\vartheta_{e,i}$. The interaction occurs during the interval $\tau = \dfrac{r_0}{\vartheta_{e,i}}$ between subsequent collisions. The probability of spinslip (of the $M$) may be described by the formulae [2]

$$P = \left(\frac{V\tau}{h}\right)^2 = \left(\frac{e^2}{r_0 \hbar^2} \cdot \frac{\vartheta_e \vartheta_i}{c^2} \cdot \frac{r_0}{\vartheta_e}\right)^2 = \left(\frac{e^2}{\hbar \vartheta_i} \cdot \frac{kT}{m_i c^2}\right)^2 = 10^{-16}$$

Coulomb cross-section of the collision is $\sigma_c = 10^{-12}$ cm$^2$. And in the result we have effective cross-section $\sigma_m = P \cdot \sigma_c = 10^{-28}$ cm$^2$. The process occurs on the free-path $l^* = \dfrac{1}{n\sigma_m} = 10^7$ cm (if $n \sim 10^{21}$); the spin-waves must be $10 \div 100$ times less than $l^*$ i.e. $\lambda_s = 10^5 \cdot 10^6$ cm. Let as find the life-time of the domain $t$: $t = \dfrac{\lambda_s}{\vartheta_i} \geq 10^5$ s (where $\vartheta_i = \dfrac{D_m}{L_h}$, $D_m$ is the coefficient of magnetical diffusion: $D_m = \dfrac{\vartheta_c \cdot r_c}{3} = 10^7$ cm; $r_c = \dfrac{m\vartheta_c c}{eB} = 1$ cm; $\vartheta_c = 10^7\ \dfrac{cm}{s}$; $L_h$ - the height of homogenous atmosphere $= 10^7$ cm) The frequency of spin wave is very low: $\omega < 10^{-4}$ sec$^{-1}$. Our observations of the Density wave and Rossby waves show the same period: $T \sim 1$ day (pictures 1-3).

On the other hand partly degenerated plasma in the photosphere may be at frontiers of the Density and Rossby waves. It is why we investigate short waves in the Solar Photosphere [1,2] and particularly logarithmic waves (the Density waves). Our pictures (1-4) show the arrangement of the tiny pores ($\varnothing = 3"\text{-}5"$) along the logarithmic spiral (1-4). In the picture la we have 15 points; the spiral has the radius $r \sim e^{k\varphi}$, ($\varphi$ -polar angle; $k = cotg\mu_0$; $\mu_0 = arccos\ M^{-1}$; $M > 1$ ($M$ is number). This case has been discussed in details in our paper [1]. In July 1989 we observed the evolution of the Density wave from $m=4$ to $m=0$ ("m" is the number of branches) during $\sim 1$ day In the pictures (la, lb) the spiral transformed into the oval with diameter $\approx 96"$ ($\approx 10^9$ cm). This shape of the configuration does not account for the well-known Rossby wave. The center of the oval has the latitude $\approx 23°$N.L. In the South hemisphere we observed the Rossby-wave during the $21^{th}\text{-}22^{th}$ July, 1989; Picture 2. Here we see $\lambda/2 = 108"$ (1 mm = 5").

There is a spiral of the shock wave in the granulation field, picture 3, (1mm = 1"). The length of the spiral arm is 25".

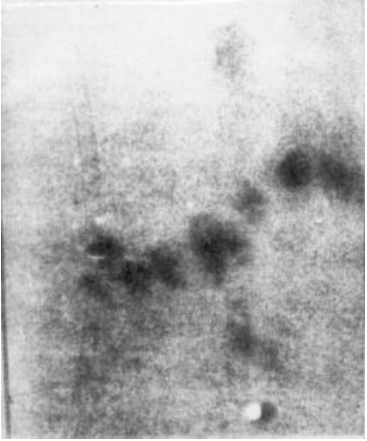

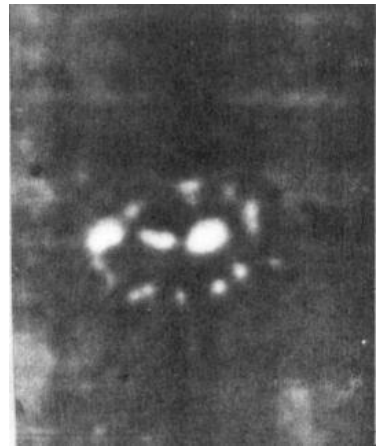

a

b

*Fig. 1.* a). Two branches of the Density wave (*m*=2); b). The last stage of the evolution of the Density wave (*m*=0).

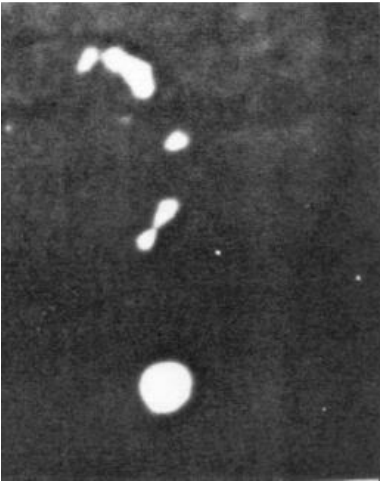

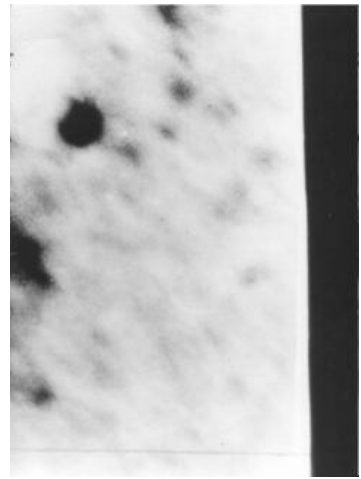

*Fig. 2.* The wave of the Density. from the film June, 15$^{th}$.

*Fig. 3.* Rossby wave (1/2 half of the wave); December 21$^{th}$ –22$^{th}$, 1989.

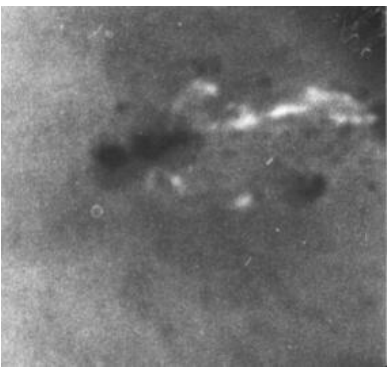

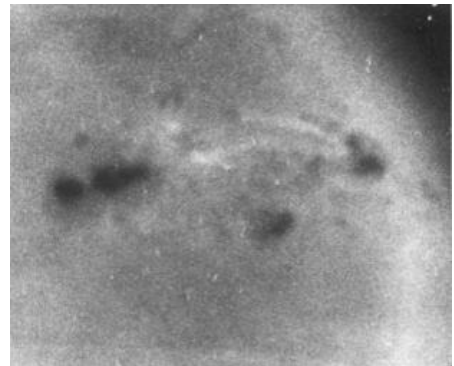

a

b

*Fig. 4.* Filtrogramm of the upper photosphere $H_\beta$ - 0,2 Å. We see the arrangement of the Moustaches of A.B. Severny along the logarithmic branches. August 7$^{th}$, 1957.

In the picture 4 we see spiral waves in the upper Photosphere ($H_\beta$ - 0",2).

These illustrations of the regular geometrical structures on the chaotic background of the Photosphere are only part of observational data. In the following paper we'll discuss the phenomena of the delicate and fragile "chains" in Photosphere. Those $\beta$ henomela are of great importance due to the fact of the emerging regular structures from irregular ones.

The author thank G.M. Seidov and A. A. Rumyantsev for the valuable advises.

[1] *M.B. Kerimbekov, A.A. Rumyantsev, E.V. Orlenko.* Doklady Academy of Science of Azerbaijan, № 1, 1990, p. 29.

[2] *M.B. Kerimbekov, A.A. Rumyantsev, E.V. Orlenko.* Doklady Academy of Science of Azerbaijan, № 2, 1990, p. 26.

[3] *M.B. Kerimbekov, A.A. Rumyantsev, E.V. Orlenko, A.A. Dovlatov.* Circular of ShAO, v. 84, 1989, p. 13.

[4] *M.B. Kerimbekov, A.A. Rumyantsev, E.V. Orlenko, A.A. Dovlatov.* Circular of ShAO, v. 85, 1989, p. 3.

[5] *E. Parker.* The cosmical magnetic fields, 1981, № 1, v. 1, Mir, Moscow, p. 71.